\documentclass[envcountsame,10pt]{llncs}

\usepackage{amsmath,amssymb,times}

\newcommand{\nat}{\mathbb{N}}
\newcommand{\dom}{{\rm dom}}
\newcommand{\dgr}{{\rm dgr}}
\newcommand{\range}{{\rm range}}
\newcommand{\rank}{{\rm rank}}

\newcommand{\Par}{{\rm Par}}
\newcommand{\lab}{{\rm lab}}
\newcommand{\pre}{{\rm pre}}
\newcommand{\edg}{{\rm edg}}
\newcommand{\true}{{\rm true}}
\newcommand{\MSO}{{\rm MSO}}
\newcommand{\GR}{{\rm GR}}

\begin{document}
\bibliographystyle{alpha}
\title{The Equivalence Problem for Deterministic\\
MSO Tree Transducers is Decidable}
\author{
  Joost Engelfriet\inst{1}\and
  Sebastian Maneth\inst{2}
\institute{
LIACS, Leiden University, The Netherlands 
\email{engelfri@liacs.nl}\and
Facult{\'e} I \& C, EPFL, Switzerland
\email{sebastian.maneth@epfl.ch}}}
\maketitle

\begin{abstract}
It is decidable for deterministic MSO definable graph-to-string
or graph-to-tree transducers whether they are equivalent on 
a context-free set of graphs. 
\end{abstract}

It is well known that the equivalence problem for nondeterministic
(one-way) finite state transducers is undecidable, 
even when they cannot read or write the empty string~\cite{gri68}.
In contrast, equivalence {\it is\/} decidable for deterministic finite state
transducers, even for two-way transducers~\cite{gur82}.
The question arises whether these results can be generalized 
from strings
to transducers working on more complex structures like, 
e.g., trees or graphs.
There is no accepted notion of finite state transducer working
on graphs; instead, it is believed that transductions expressed in 
monadic second-order logic (MSO) are the natural counterpart of 
finite state transductions on graphs.
The idea is to define an output graph by interpreting fixed MSO
formulas on a given input graph.
In fact, if the input and output graphs of such an MSO graph transducer are
strings, then the resulting transductions (in the deterministic case)
are precisely the deterministic 
two-way finite state transductions~\cite{enghoo01}.
Hence, by the above, equivalence is decidable for deterministic
MSO string transducers.
A nondeterministic MSO graph transducer can easily simulate
a nondeterministic finite state transducer 
that cannot read the empty string; hence, equivalence
is undecidable.
Actually, even for deterministic MSO graph transducers 
equivalence is undecidable.
This is due to the fact that MSO is undecidable
for graphs
(Propositions~5.21 and 5.2.2 of~\cite{cou97_short}).
The question remains whether deterministic MSO tree transducers
have a decidable equivalence problem.
Recently, these transducers have been characterized by
certain attribute grammars~\cite{bloeng00} and 
macro tree transducers~\cite{engman99a}.
However, for both models it is unknown whether equivalence 
is decidable.
Here we give an affirmative answer: equivalence of deterministic
MSO tree transducers is decidable.
This result has several applications; for instance, 
it implies that XML queries of linear size increase have
decidable equivalence, by the results 
of~\cite{milsucvia03},~\cite{engman03b},~\cite{engman03a},
and~\cite{man03c}.
Our proof generalizes the one of~\cite{gur82} (see also~\cite{iba82}): 
it is based on the
fact that certain sets are semilinear.
The reader is assumed to be familiar with MSO on graphs and
with MSO graph transducers, see, e.g.,~\cite{cou97_short,cou94a}.

{\bf Convention:} 
All lemmas stated in this paper are {\it effective}.

A graph alphabet is a pair $(\Sigma,\Gamma)$ of
alphabets of node and edge labels, respectively.
A graph over $(\Sigma,\Gamma)$ is a tuple $(V,E,\lambda)$ where
$V$ is the finite set of nodes, 
$E\subseteq V\times\Gamma\times V$ is the set of edges,
and $\lambda: V\to\Sigma$ is the node labeling function.
The set of all graphs over $(\Sigma,\Gamma)$ is
denoted $\GR(\Sigma,\Gamma)$.
The language $\MSO(\Sigma,\Gamma)$ of 
monadic second-order (MSO) formulas over $(\Sigma,\Gamma)$
uses node variables $x,y,\dots$ and node-set variables
$X,Y,\dots$; both can be quantified with $\exists$ and $\forall$. 
It has atomic formulas 
$\lab_\sigma(x)$ for $\sigma\in\Sigma$, denoting that $x$ is labeled
$\sigma$,
$\edg_\gamma(x,y)$ for $\gamma\in\Gamma$, denoting that
there is a $\gamma$-labeled edge from $x$ to $y$,
and $x\in X$ denoting that $x$ is in $X$.
For $g\in\GR(\Sigma,\Gamma)$ and a closed formula 
$\psi$ in $\MSO(\Sigma,\Gamma)$
we write $g\models\psi$ if $g$ satisfies $\psi$;
similarly, if $\psi$ has free variables $x$ or $x,y$ and
$u,v$ are nodes of $g$, then we write $(g,u)\models\psi$ 
or $(g,u,v)\models\psi$ if $g$ satisfies $\psi$ with $x=u$ or
with $x=u$, $y=v$, respectively.

Let $(\Sigma_1,\Gamma_1),(\Sigma_2,\Gamma_2)$ be graph
alphabets.
A {\it deterministic MSO graph transducer\/} $M$ 
({\it from\/} $(\Sigma_1,\Gamma_1)$ {\it to\/} $(\Sigma_2,\Gamma_2)$)
is a tuple
$(C,\varphi_{\text{dom}},\Psi,X)$ where
$C$ is a finite set of {\it copy names},
$\varphi_{\text{dom}}\in\text{MSO}(\Sigma_1,\Gamma_1)$ is
the closed {\it domain formula},
$\Psi=\{\psi_{c,\sigma}(x)\}_{c\in C, \sigma\in\Sigma_2}$ is a 
family of {\it node formulas}, i.e., 
MSO formulas $\psi_{c,\sigma}(x)$
over $(\Sigma_1,\Gamma_1)$ with one free variable $x$,
and
$X=\{\chi_{c,c',\gamma}(x,y)\}_{c,c'\in C,\gamma\in\Gamma_2}$
is a family of {\it edge formulas}, i.e.,
MSO formulas $\chi_{c,c',\gamma}(x,y)$ 
over $(\Sigma_1,\Gamma_1)$
with two free variables $x,y$.

Given $g\in\GR(\Sigma_1,\Gamma_1)$, 
the graph $h=\tau_M(g)\in \GR(\Sigma_2,\Gamma_2)$ is defined if 
$g\models\varphi_{\text{dom}}$, and then
$V_h=\{(c,u)\mid c\in C, u\in V_g$, there is exactly one
$\sigma\in\Sigma_2$ such that $(g,u)\models\psi_{c,\sigma}(x)\}$,
$E_h=\{((c,u),\gamma,(c',u'))\mid (c,u),(c',u')\in V_h,
\gamma\in\Gamma_2$, and $(g,u,u')\models\chi_{c,c',\gamma}(x,y)\}$,
and
$\lambda_h=\{((c,u),\sigma)\mid(c,u)\in V_h,\sigma\in\Sigma_2$, and
$(g,u)\models\psi_{c,\sigma}(x)\}$.
Hence, $\tau_M$ is a partial function from 
$\GR(\Sigma_1,\Gamma_1)$ to $\GR(\Sigma_2,\Gamma_2)$ with
$\dom(\tau_M)= 
\{g\in\GR(\Sigma_1,\Gamma_1)\mid g\models\varphi_{\text{dom}}\}$.

A ({\it nondeterministic\/}) {\it MSO graph transducer\/} is obtained from a
deterministic one by allowing all formulas to use fixed 
free node-set variables $Y_1,Y_2,\dots$, called parameters.
For each valuation of the parameters (by sets of nodes of the input
graph) that satisfies the domain formula, the other formulas define
the output graph as before. Hence each such valuation may lead to a 
different output graph for the given input graph.
Thus, $\tau_M\subseteq\GR(\Sigma_1,\Gamma_1)\times\GR(\Sigma_2,\Gamma_2)$.

The following lemma contains a basic fact about MSO definable
graph transductions; see, e.g., Proposition~3.2 in~\cite{cou94a}.

\begin{lemma}\label{lm_mso}
The (deterministic) MSO graph transductions are closed under composition.
\end{lemma}

{\bf Notation.} Let $M_1 ; M_2$ denote a transducer $M$
for which $\tau_M = \tau_{M_2}\circ \tau_{M_1}$;
note that $M$ is deterministic, if $M_1$ and $M_2$ are.
By Lemma~\ref{lm_mso}, $M_1 ; M_2$ effectively exists.

In the sequel we often identify a transducer $M$ with its 
transduction $\tau_M$, and simply write, e.g., $M(g)$ in place
of $\tau_M(g)$. 

Let $M$ be an MSO graph transducer and let $X,Y$ be sets of graphs.
Then $M$ is called an MSO $X$-to-$Y$ transducer, if 
$\dom(M)\subseteq X$ and $\range(M)\subseteq Y$, and
it is an MSO $X$ transducer if additionally $Y=X$.

A {\it discrete graph\/} ({\it dgraph}, for short) is a graph
without edges.
Let $g$ be a dgraph over $(\Sigma,\varnothing)$ with
$\Sigma=\{\sigma_1,\dots,\sigma_k\}$.
Define $\Par(g)$ as the vector $(n_1,\dots,n_k)$ in $\nat^k$
such that, for $1\leq i\leq k$, 
$n_i$ is the number of $\sigma_i$-labeled nodes in $g$.
Similarly, for a string $w\in\Sigma^*$, $\Par(w)$ 
is the vector in $\nat^k$ such that the $i$-th component
is the number of $\sigma_i$'s in $w$.
We denote by $\dgr(w)$ the (unique) dgraph $g$ such that
$\Par(g)=\Par(w)$.
For a set $S$ of dgraphs or strings,  
$\Par(S)$ is the set of all $\Par(g)$ for $g\in S$.
A set $P\subseteq\nat^k$ is {\it semilinear\/} if there exists
a regular language $R$ such that $P=\Par(R)$.
The set $S$ is {\it Parikh\/} if $\Par(S)$ is semilinear.
Note that since $\Par(R)=\emptyset$ iff $R=\emptyset$, 
emptiness of semilinear sets is decidable. 

A set of graphs is {\it NR\/} if it is generated by a
context-free \underline{n}ode \underline{r}eplacement graph grammar, see,
e.g.,~\cite{eng97,cou94a}; it is also called C-edNCE or VR. 

\begin{lemma}(Theorem~7.1 of~\cite{cou94a})\label{lm_Parikh}
The images of NR sets of graphs under 
MSO graph-to-dgraph transductions are Parikh.
\end{lemma}

In fact, the class of NR sets of graphs is closed under 
MSO graph transductions (see Theorem~4.2(3) of~\cite{cou94a}, or
Section~5 of~\cite{eng97}) and NR sets of graphs are
Parikh (see Proposition~4.11 of~\cite{eng97}).

A useful property of semilinear sets is their (effective) closure
under intersection. It implies the following lemma.

\begin{lemma}\label{lm_sem}
It is decidable for a semilinear set $S\subseteq\nat^2$ whether there 
exists an $n\in\nat$ such that $(n,n)\in S$.
\end{lemma}
\begin{proof}
Let 
$P=\{(n,n)\mid n\in\nat\}=\Par((ab)^*)$.
The lemma holds because
$S\cap P$ is semilinear~\cite{ginspa64} and semilinear sets have a decidable
emptiness problem.
\qed
\end{proof}

We identify the
string $w=a_1a_2\cdots a_n$ with the graph that has
$\#$-labeled nodes $v_1,\dots,v_{n+1}$ and,
for $1\leq i\leq n$,
an $a_i$-labeled edge from $v_i$ to $v_{i+1}$.
For $1\leq i\leq n$,
we denote by $w/i$ the $i$-th letter $a_i$ of $w$.

\begin{lemma}\label{lm_comstr}
Let $\Delta$ be an alphabet and $a\in\Delta$.
There exists an MSO string-to-dgraph
transducer $N_\Delta^a$ such
that for every $w\in\Delta^*$,
\[
N_\Delta^a(w)=\{\dgr(a^n)\mid
w/n=a\}.
\]
\end{lemma}
\begin{proof}
The transducer $N_\Delta^a$ uses one parameter $Y_1$ to 
nondeterministically choose a node $v$ 
that has an outgoing $a$-labeled edge.
It copies $v$ and all input nodes to the left of $v$, 
and labels them $a$.
There are no edge formulas because dgraphs have no edges.
Define $N_\Delta^a=(\{1\},\varphi_{\text{dom}}(Y_1), 
\psi_{1,a}(x,Y_1),\varnothing)$
with 
\[
\begin{array}{lcl}
\varphi_{\text{dom}}(Y_1)&\equiv &
\text{singleton}(Y_1)\wedge
(\exists x)(\exists y)(\edg_a(x,y)\wedge x\in Y_1)\\
\psi_{1,a}(x,Y_1)&\equiv&(\exists y)(x\preceq y\wedge y\in Y_1)
\end{array}
\]
where $\text{singleton}(Y_1)$
expresses that $Y_1$ is a singleton, 
and $x\preceq y$ that there is a path from $x$ to $y$.
\qed
\end{proof}

We denote the disjoint union of graphs $h_1$ and $h_2$ 
by $h_1\uplus h_2$. 

\begin{lemma}\label{lm_union}
Let $M_1,M_2$ be MSO graph transducers.
There exists an MSO graph transducer $M$, denoted
$M_1\uplus M_2$, such that for every graph $g$, 
\[
M(g)=\{ h_1\uplus h_2\mid h_1\in M_1(g),
h_2\in M_2(g)\}.
\]
\end{lemma}
\begin{proof}
Let $M_1=(C_1,\varphi_1,\Psi_1,X_1)$ and 
$M_2=(C_2,\varphi_2,\Psi_2,X_2)$.
We may assume w.l.o.g. that $C_1$ is disjoint from $C_2$
and that the parameters of $M_1$ are disjoint from
those of $M_2$.
Then $M=(C_1\cup C_2,\varphi_1\wedge\varphi_2, \Psi_1\cup\Psi_2, X_1\cup
X_2\cup X)$
realizes the desired transduction, where
all edge formulas in $X$ are set to false.
\qed
\end{proof}

\begin{lemma}\label{lm_M}
Let $M_1,M_2$ be MSO graph-to-string transducers 
and let $a,b$ be distinct symbols.
There exists an MSO graph-to-dgraph transducer 
$M^{a,b}$ such that for every graph $g$,
\[
M^{a,b}(g)=\{\dgr(a^mb^n)\mid 
\exists h_1\in M_1(g), h_2\in M_2(g):
h_1/m=a\text{ and }h_2/n=b\}.
\]
\end{lemma}
\begin{proof}
Let $M_i$ be from $(\Sigma_i,\Gamma_i)$ to $(\{\#\},\Delta_i)$ 
for $i\in\{1,2\}$.
If $a\not\in\Delta_1$ or $b\not\in\Delta_2$ then 
let $M^{a,b}=(\varnothing,\text{false},\varnothing,\varnothing)$.
Otherwise
define 
$M^{a,b}= (M_1 ; N^a_{\Delta_1})\uplus (M_2 ; N^b_{\Delta_2})$
according to Lemmas~\ref{lm_mso}, \ref{lm_comstr}, and~\ref{lm_union}.
\qed
\end{proof}

Let $\Sigma$ be a ranked alphabet, i.e., an alphabet $\Sigma$
together with a mapping $\rank_\Sigma:\Sigma\to\nat$.
Let $m$ be the maximal rank of symbols in $\Sigma$.
A {\it tree\/} (\text{\it over $\Sigma$\/}) 
is an acyclic, connected graph in
$\GR(\Sigma,\{1,\dots,m\})$,
with exactly one node that has no incoming edges (the root),
and, for $\sigma\in\Sigma$, every $\sigma$-labeled node
has exactly $\rank_\Sigma(\sigma)$ outgoing edges,
labeled $1,2,\dots,\rank_\Sigma(\sigma)$, respectively.

For a relation $R\subseteq A\times B$ and a set $D\subseteq A$, denote by
$R|_D$ the restriction of $R$ to $D$, i.e.,
$R|_D = \{ (a,b)\in R\mid a\in D\}$.

\begin{theorem}\label{theo_main}
It is decidable for deterministic MSO graph-to-string or 
graph-to-tree transducers 
$M_1,M_2$ and an NR set $D$ of graphs whether 
$\tau_{M_1}|_D=\tau_{M_2}|_D$.
\end{theorem}
\begin{proof}
We start with the graph-to-string case.
For $i\in\{1,2\}$ let $D_i=\dom(M_i)\cap D$.
We first show that it is decidable whether $D_1=D_2$. 
Clearly, $D_1=D_2$ if and only if $\Par(E(D))=\emptyset$, 
where $E$ is the deterministic MSO graph-to-dgraph transducer that 
removes the edges of all graphs in the symmetric 
difference of $\dom(M_1)$ and $\dom(M_2)$: 
$E= (\{1\}, \neg(\varphi_1 \leftrightarrow \varphi_2), 
\{\psi_{1,\sigma}(x)\}_{\sigma\in\Sigma}, \emptyset\}$ 
where $\varphi_i$ is the domain formula of $M_i$ for $i\in\{1,2\}$, 
$\Sigma$ is the node alphabet of $D$, 
and $\psi_{1,\sigma}(x)= \lab_\sigma(x)$ for $\sigma\in\Sigma$.
By Lemma~\ref{lm_Parikh}, $\Par(E(D))$ is effectively semilinear,
and hence its emptiness can be decided. 
If $D_1\not=D_2$ then we are finished and know that
$\tau_{M_1}|_D\not=\tau_{M_2}|_D$. Assume now that $D_1=D_2$.

Let $M_i$ have output edge alphabet $\Delta_i$,
for $i\in\{1,2\}$, and
let $\$$ be a symbol not in $\Delta= \Delta_1\cup\Delta_2$.
We define deterministic MSO graph-to-string transducers
$M_i^\$ = M_i ; N$ such that 
$M_i^\$(g)=M_i(g)\$$ for all $g\in\dom(M_i)$. 
Here $N$ is the deterministic MSO string transducer
$(C,
\text{true},
\{\psi_{1,\#}(x),\psi_{2,\#}(x)\},
\{\chi_{c,c',\delta}(x,y)\}_{c,c'\in C,\delta\in\Delta\cup\{\$\}})$
such that
$C=\{1,2\}$,
$\psi_{1,\#}(x)\equiv\text{true}$,
$\psi_{2,\#}(x)\equiv
\chi_{1,2,\$}(x,y)\equiv
\neg(\exists z)\bigvee_{\delta\in\Delta}\edg_\delta(x,z)$
and, for $\delta\in\Delta$, 
$\chi_{1,1,\delta}(x,y)\equiv\edg_\delta(x,y)$;
all other edge formulas are set to false.

Since now all output strings end on the special marker $\$$,
$\tau_{M_1}|_D \not= \tau_{M_2}|_D$ iff
\[
\exists a\exists b: (d(a,b)
\wedge \exists n\exists g:
(g\in D_1\wedge
M_1^\$(g)/n=a\ \wedge\ M_2^\$(g)/n=b))
\]
where $d(a,b)$ denotes the statement
$a,b\in(\Delta\cup\{\$\})\wedge a\not=b$.
For given $a,b$, let $M^{a,b}$ be the transducer of
Lemma~\ref{lm_M} for $a,b,M_1^\$,M_2^\$$.
Then the statement displayed above holds if and only if
\[
\begin{array}{ll}
&\exists a\exists b: (d(a,b) \wedge \exists n:
\dgr(a^nb^n)\in M^{a,b}(D)))\\
\text{iff}&
\exists a\exists b: (d(a,b) \wedge 
\underbrace{
\exists n: (n,n)\in\Par(M^{a,b}(D))) }_{P(a,b)} )
\end{array}
\]
By Lemma~\ref{lm_Parikh},
$\Par(M^{a,b}(D))$ is effectively semilinear.
By Lemma~\ref{lm_sem} this means that $P(a,b)$ is decidable.
Since there are only finitely many $a,b$ with
$d(a,b)$, the statement is decidable.

We now reduce the graph-to-tree case to the graph-to-string case.
Let $\Delta$ be a ranked alphabet and let $m$ be the maximal rank
of its elements.
There is a deterministic 
MSO tree-to-string transducer $M_\Delta$ that 
translates every tree $t$ over $\Delta$ into the string 
$\pre(t)$ of its node labels in pre-order.
Clearly, if we associate with a deterministic
MSO graph-to-tree transducer $M$ 
(from $(\Sigma,\Gamma)$ to $(\Delta,\{1,\dots,m\})$)
the deterministic MSO graph-to-string transducer
$\widehat{M}=M ; M_\Delta$, then 
$M_1$ is equivalent to $M_2$ on $D$ 
if and only if 
$\widehat{M_1}$ is equivalent to $\widehat{M_2}$ on $D$.
Let $M_\Delta=(\{1,2\},\true,\{\psi_{1,\#},\psi_{2,\#}\},
\{\chi_{c,c',\delta}\}_{c,c'\in \{1,2\},\delta\in\Delta})$
with
$\psi_{1,\#}\equiv\true$,
$\psi_{2,\#}\equiv\text{root}(x)$, where $\text{root}(x)$ expresses
that $x$ is the root node.
Further, for $\delta\in\Delta$,
$\chi_{1,1,\delta}\equiv
\lab_\delta(x)\wedge \pi(x,y)$
and
$\chi_{1,2,\delta}\equiv \lab_\delta(x)\wedge\text{root}(y)\wedge
\neg(\exists z)\,\pi(x,z)$
where 
$\pi(x,y)$ expresses that $y$ is the successor of $x$ in the pre-order.
\qed
\end{proof}

\subsubsection*{String and Tree Transductions}

Clearly, Theorem~\ref{theo_main} also holds if we restrict
the input graphs to strings or trees. 
In particular, deterministic MSO $X$-to-$Y$ transducers have 
decidable equivalence for all $X,Y\in\{\text{string},\text{tree}\}$. 
For string transducers this reproves the decidability result of 
\cite{gur82} (through \cite{enghoo01}).
For trees we obtain the following new decidability result.

\begin{corollary}
The equivalence problem is decidable for deterministic MSO tree transducers.
\end{corollary}

Of course, even stronger statements hold; namely,
given an NR set $D$ of strings or trees, it is decidable if 
two deterministic MSO $X$-to-$Y$
transducers are equivalent when restricted to $D$.
For string transducers this means the following. 

\begin{corollary}\label{cor_nr}
It is decidable whether two deterministic two-way finite state transducers 
are equivalent on an NR set of strings.
\end{corollary}

As discussed in Section~6 of~\cite{eng97}, 
the NR sets of strings are the same as the ranges of
deterministic tree-walking tree-to-string transducers. 
They properly contain, for instance, the context-free languages
and the ranges of deterministic two-way finite state transducers. 
Since the NR sets of strings form a full AFL of Parikh languages,
Corollary~\ref{cor_nr} is in fact 
a special case of the general decidability result for deterministic 
two-way finite state transducers in Theorem~5 of~\cite{iba82}.
It is incomparable
to the decidability of equivalence of two such transducers on an 
NPDT0L language~\cite{culkar87}.

The two statements of the next corollary 
follow from the characterizations of deterministic MSO
definable tree transductions in~\cite{bloeng00} and~\cite{engman03a},
respectively. 
Note that a tree transducer is of linear size increase 
if the size of the output tree is 
at most linear in the size of the input tree. 

\begin{corollary}
The equivalence problem is decidable \\
(1) for single-use restricted attributed tree transducers and \\
(2) for deterministic macro tree transducers of linear size increase.
\end{corollary}

This result is incomparable with the decidability of 
the equivalence problem for nonnested separated attributed/macro 
tree transducers proved in~\cite{coufra82}.
It remains open whether the equivalence problem is decidable
for attributed tree transducers and 
for deterministic macro tree transducers.

In~\cite{milsucvia03} the $k$-pebble tree transducer was introduced,
and claimed to subsume (the tree translation core of) all known
XML query languages. Hence, we call deterministic pebble tree
transducers
{\it deterministic XML queries}. Such queries can be simulated by
compositions of macro tree transducers~\cite{engman03b}.
If such compositions are of linear size increase, then they are
MSO definable~\cite{man03c}.

\begin{corollary}
The equivalence problem is decidable for deterministic 
XML queries of linear size increase.
\end{corollary}

\end{document}